# Estimation of age-specific excess mortality of men and women with rheumatoid arthritis (RA) in Germany

Ralph Brinks, Chair for Medical Biometry and Epidemiology, Witten/Herdecke University, Faculty of Health/School of Medicine, D-58448 Witten, Germany

## Abstract

A MCMC approach is used to estimate the age-specific mortality rate ratio for German men and women with RA. For constructing priors, we calculate a range of admissible values from prevalence and incidence data based on about 60 million people in Germany. Using these priors, MCMC mimics and compares estimated mortality to the findings of a recent register study from Denmark. It is estimated that the mortality rate ratio is highest in the young ages (4.0 and 3.5 for men and women aged 17.5 years, respectively) and declines towards higher ages (1.0 and 1.2 for men and women aged 92.5 years, respectively). The lengths of the credibility intervals decrease from younger towards older ages.

## Introduction

Aggregated prevalence and incidence data for chronic conditions becomes more and more available. In principle, it is possible to estimate the excess mortality of people with a chronic condition compared to those without the condition from age-specific prevalence and incidence data [Bri19]. The idea is based on the illness-death model for chronic conditions (see e.g., [Mei19] and references therein) and uses the fact, that prevalence, incidence and excess mortality are related by a partial differential equation (PDE) [Bri15]. For ages above 50 years, this has successfully been accomplished in the field of type 2 diabetes [Tön18], however, for younger age groups, the method is instable with respect to inaccuracies in case detection. False positive and false negative cases in prevalence or incidence data can impose huge biases in estimates of the excess mortality [Bri21]. By false positive and false negative cases, we refer to two types of error that may occur in the aggregated data set: People with the chronic condition in reality, on the one hand, might not have the diagnosis coded in the aggregated data and can hence be assumed to be false negatively coded. On the other hand, people without the chronic condition in reality might have a corresponding diagnosis in the data set. Henceforth, we refer the later as false positive findings in the data set. By opposing the diagnoses coded in the data set with "reality", i.e. the "gold standard", such as a medical diagnosis of a specialist physician, the diagnoses underlying the aggregated data can be interpreted similar to a diagnostic test. As in diagnostic tests, henceforth, we denote the percentage of false positive diagnoses in the data set as false positive ratio (*FPR*). Similarly, the percentage of true positive diagnoses are denoted as sensitivity (*se*).

In this paper, we propose a Bayesian approach to overcome the problems with potential inaccuracies in case detection.

## Methods

Based on the illness-death model for chronic conditions, in [Bri15] we could derive a PDE that relates the temporal change of the age-specific prevalence $p = p(t, a)$, i.e. the proportion of people aged $a$ at calendar time $t$ with the chronic condition with the incidence rate $i(t, a)$, mortality rate of people with and without disease, $m_1(t, a)$ and $m_0(t, a)$, respectively. The PDE is given in Equation (1).

$$(\partial_t + \partial_a)\, p = (1 - p)\, \{i - p \times (m_1 - m_0)\} \qquad (1)$$

We express excess mortality in terms of the age-specific mortality rate ratio ($R$) which is the quotient $R(t, a) = m_1(t, a)/m_0(t, a)$. In this case, the PDE reads as

$$(\partial_t + \partial_a)\, p = (1 - p)\, \{i - m \times p\, (R - 1) / [1 + p\, (R - 1)]\}, \qquad (2)$$

where $m(t, a)$ is the general mortality rate in the overall population (i.e., irrespective whether with or without the chronic condition). For given $(\partial_t + \partial_a)\, p = \partial p$, $p$, $i$ and $m$, Equation (2) can be solved for $R$:

$$R = 1 + \{i \times (1 - p) - \partial p\}/\{p \times [(m - i) \times (1 - p) + \partial p]\} \qquad (3)$$

The general mortality $m$ of the overall population has been obtained from the Federal Statistical Office of Germany.

Equations (1), (2) and (3) hold true for the true prevalence $p$ and incidence rate $i$. If we assume an observed prevalence $p^{(obs)}$ and an observed incidence $i^{(obs)}$ in the aggregated data set (possibly imperfect with respect to diagnostic accuracy), the true prevalence and incidence can be obtained from in terms of Equations (4a) and (4b) using the *se* and *FPR*.

$$p = (p^{(obs)} - FPR_p)/(se_p - FPR_p) \qquad (4a)$$

and

$$i = (i^{(obs)} - FPR_i)/(se_i - FPR_i). \qquad (4b)$$

In Equations (4a) and (4b), *se* and *FPR* in the age-specific prevalence and incidence (indicated by the sub-indices) do not have to be the same.

For estimation of $R$, we use an approach based on two steps. In the first step, we estimate a range of admissible $R$ values using the PDE. Similar to [Tön18], we use aggregated prevalence and incidence data about rheumatoid arthritis (RA) comprising about 60 million people aged ≥15 years from the German statutory health insurance (SHI) [Ste17]. These data are provided from the Zentralinstitut der Kassenärztlichen Versorgung (Zi), which collects all outpatient diagnoses of insurants of the German SHI. The derived information about $R$ is used as weakly-informative prior for the Markov chain Monte Carlo (MCMC) algorithm. In the second step, we mimic a study similar to a register study accomplished recently in Denmark [Lop19]. With the MCMC algorithm, we simulate samples from the distribution of the unknown $R$ values given the data from the Danish study.

Assumptions A for calculation of admissible $R$ values from the Zi data:
1) For a specific age $a$, sensitivity (*se*) with respect to incidence ($se_i$) and prevalence ($se_p$) are the same: $se_i(a) = se_p(a) =: se(a)$.
2) There is no change of $se(a)$ over the study period 2009 to 2015. In other words, *se* does not depend on calendar time $t$ during the study period.
3) The same as in assumptions 1) and 2) hold true for the false positive ratio (*FPR*): $FPR_i(a) = FPR_p(a) =: FPR(a)$. The assumption of same *se* and *FPR* with respect to prevalence and incidence is justified, because prevalent and incident cases in the Zi data are obtained from ascertained diagnoses of all physicians treating the insurants of the SHI. If prevalence data suffer from false positive or false negative findings, incidence data will have the same problem to the same extent.
4) Age-specific incidence does not depend on calendar time $t$: $i(t, a) = i(a)$ for all ages $a$.

Algorithm 1: calculation of admissible $R$ values from Zi data:
1) For each age $a$: Draw random sample of sensitivity values $se(a)$ in the range from 70 to 100 percent.
2) For each age $a$: Draw random sample of false positive ratios $FPR(a)$ values in the range from 0 to the minimum of $p^{(obs)}(2009, a)$, $p^{(obs)}(2015, a)$, $i^{(obs)}(a)$. The minimum is chosen to guarantee that the unknown true values $p(2009, a)$, $p(2015, a)$, $i(a)$ given by Equations (4a) and (4b) are non-negative.
3) For each pair ($se$, $FPR$) from steps 1) and 2): Calculate the true prevalence $p(2009, a)$, $p(2015, a)$, $i(a)$ via Eqs. (4a), (4b) and calculate $R(a)$ via Eq. (3).
4) Delete negative $R(a)$ values and calculate the range of the remaining $R(a)$ values.

Assumptions B for estimation of $R$ values from the Danish register data:
1) The age-specific sensitivity and specificity of the Danish register is the same as in the Zi data.
2) The age-specific $R$ values observed in Denmark are those values that would have been observed in Germany. This assumption is justified by an argument of Breslow that $R$ values are stable across different populations [Bre80, p. 59].
3) The graph of the age-specific logarithmized $R$ values, $\log(R(a))$ over age $a$ is "close" to a straight line. This assumption is justified by observations from other chronic diseases [Car20].

Algorithm 2: MCMC estimation of R values from the Danish register data:
1) For each age $a$: Draw from uniform distribution from the whole range of admissible $R(a)$ (see step 4 in Algorithm 1.
2) Use prior assumption and penalize choices where the graph of $\log(R(a))$ over age $a$ are non-straight lines. Deviation from the straight line assumption is assessed in terms of squared Pearson's correlation coefficient.
3) Use $p^{(Obs)}$ to estimate total hazard ratio ($HR$) that would have been observed if the age and sex distribution was like in the Danish study. Penalize using the likelihood of estimated $HR$ value based on the $R$ values compared to the observed $HR$ from the Danish study.

We use an MCMC algorithm with chain length of 300,000 with burn-in of 40% of the chain length.

The source code for use with the statistical software R (The R Foundation for Statistical Computing) including the data is available in the public access online repository under digital object identifier (DOI) 10.5281/zenodo.7651403.

# Results

Figure 1 shows the estimated age-specific mortality rate ratios $R$ with 95% credibility intervals for men (left) and women (right). Credibility intervals (CI) are computed as empirical 2.5% and 97.5% quantiles of the estimated values for the MCMC ignoring the values from burn-in. The median estimates for each age are depicted as small horizontal lines on the vertical lines indicating the 95% CI. For both sexes, we see a downward trend of the median estimates over age. Until age of 40 median $R$ values for men are slightly above those of women. After 40 years of age, this relation reverses. The size of the credibility intervals decreases with age, which is consistent with the observation in [Bri21] that in lower age groups estimation of mortality rate ratios is prone to uncertainty in sensitivity and specificity of the underlying data about diagnoses of RA.

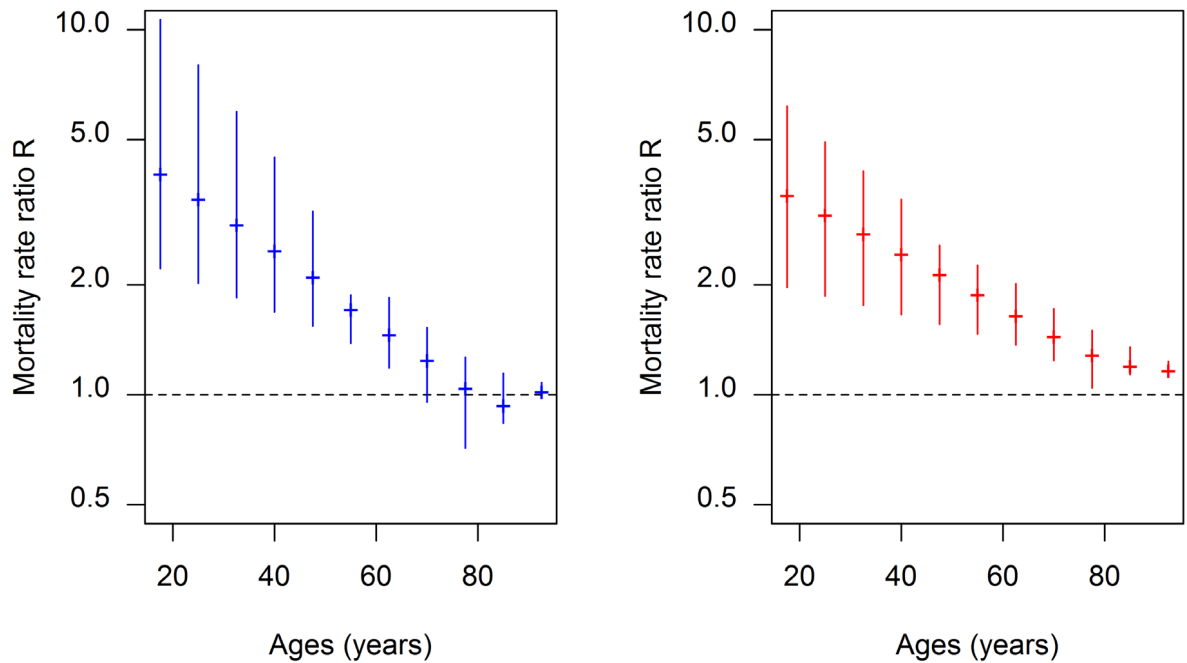

**Figure 1: Age specific mortality rate ratio *R* for German men (left) and women (right) with rheumatoid arthritis.**

# Discussion

For the first time in Germany, we were able to estimate the age-specific mortality rate ratio *R* for male and female patients with RA compared to those without RA on a population level. Instead of directly examining death cases (e.g, by running a disease register and collecting death cases), we use an indirect estimation method based on an MCMC approach. First, we sample prior distributions obtained from prevalence and incidence data about RA from 60 million people from the statutory health insurance in Germany. Prevalence and incidence data of a chronic condition in principle are sufficient to estimate the mortality rate ratio via Eq. (3) [Tön18]. However, in younger age-groups the method is prone to false positive and false negative cases, which requires additional considerations. Here, we decided to use the priors based on the prevalence and incidence in combination with data from a Danish register study. Denmark and Germany are similar with respect to population structure and health services, which gave rise to the idea of mimicking the Danish study with the prior information obtained from Eq. (3) and German prevalence data. In the MCMC approach, given the candidate *R* values from the priors, it is estimated which overall hazard ratio would have been observed if the age- and sex-distribution would have been like in the Danish study. Differences are penalized via a likelihood function.

Advantages of our approach lies in the fact that mortality data by disease status are not necessary. Typically, these data are obtained by disease registers or cohort studies, which might be costly. In our approach, we need prevalence and incidence data about RA and and mortality rates of the overall German population. The later have been obtained from the Federal Statistical Office.

As a drawback of our approach, we need the assumption that the mortality rate ratio of patients with diagnosed RA in Denmark are the same as in Germany. Apart from similar treatment of patients with RA in Denmark and Germany, we also assume that the case detection of people with RA are similar in Denmark and Germany.

# Contact

Ralph Brinks
Chair for Medical Biometry and Epidemiology
Witten/Herdecke University
Faculty of Health/School of Medicine
D-58448 Witten
Germany
Ralph.Brinks@uni-wh.de